\documentclass[12pt]{article} 
\usepackage{geometry}
\usepackage{caption}
\usepackage{array}

\usepackage{chngcntr}
\counterwithout{equation}{section}

\newcommand{\Team}{2322645} % Replace 1111111 with your Contest Team Control Number
\usepackage{wrapfig}

\usepackage{csvsimple}%for importing the species data from csv
\usepackage{enumitem}%for making lists denser
\usepackage{amsmath,amssymb,amsthm,amsfonts}

\usepackage{hyperref}
\hypersetup{
    colorlinks = true,
    citecolor=black,
    filecolor=black,
    linkcolor=black,
    urlcolor=blue,
    }
\usepackage{lastpage}

\usepackage{booktabs} 

\usepackage{siunitx}

\usepackage{adjustbox}

\usepackage{xcolor}
\usepackage{fancyhdr}
\usepackage{graphicx} % load after xcolor

\allowdisplaybreaks

\usepackage{changepage}

\usepackage[vlined,ruled,linesnumbered]{algorithm2e}
\usepackage{tikz}
\usetikzlibrary{decorations.pathmorphing}

\title{How many Wordle words will Wordle guessers guess if Wordle's Wednesday Wordle word is ``Eerie''?}

\date{\vspace{-5ex}}
\usepackage{biblatex}
\author{Steven DiSilvio, Anthony Ozerov, Leon Zhou}
\date{February 20, 2023}

\begin{document}

\graphicspath{{.}}  % Place your graphic files in the same directory as your main document
\DeclareGraphicsExtensions{.pdf, .jpg, .tif, .png}

%%%%%%%%%%% Begin Summary %%%%%%%%%%%

\maketitle

\begin{abstract}
    With the rising popularity of Wordle, people have eagerly taken to Twitter to report their results daily by the tens of thousands. In this paper, we develop a comprehensive model which uses this data to predict Wordle player performance and reporting given any word and date. To do so, we first decompose words into quantifiable traits associated with relevant difficulty characteristics. Most notably, we formulate a novel Wordle-specific entropy measure which quantifies the average amount of information revealed by typical players after initial guesses. We also develop a method to represent the distribution of player attempts, and hence the observed difficulty of a word, using just two values corresponding to the cumulative mass function of the Beta distribution.  Consequently, we are able to use a preliminary Lasso regression to isolate the most relevant predictors of word difficulty, which we then use in a Bayesian model. 
    
    For a given word and date, our Bayesian model predicts the distribution of the number of guesses by players (i.e. the reported player performance), the number of player reports, and the number of players reporting playing in hard-mode. To accomplish these three tasks, it is made up of three submodels which are conditionally independent given the data, making it efficient to sample from its posterior using Markov Chain Monte Carlo (MCMC). Our model is able to predict outcomes for new data and retrodict for old data, and we empirically demonstrate the success of our model through retrodictions on unseen data. Most notably, our model does not just provide such simple point estimates and prediction intervals, but full posterior distributions.
\end{abstract}

\newpage

\tableofcontents 
%\newpage
%%%%%%%%%%%%%%%%%%%%%%%%%%%%%%

% \section{Place to throw ideas}

% Date
% \begin{itemize}
%     \item Day of Week
%     \item Growth Rate
%     \item Special Dates
% \end{itemize}

% Word
% \begin{itemize}
%     \item Vowel: \#
%     \item Word Order: [\#,\#,\#,\#,\#]
%     \item Letter Freq: [\#,\#,\#,\#,\#]
%     \item Unique Letters: \#
%     \item Usage: \#
%     \item Entropy: \#
% \end{itemize}

\newpage

\section{Introduction}
Wordle is a language-based game currently owned by the New York Times that became a viral sensation in early 2022. The goal of the game is simple to understand: At the the start of each day there is a 5-letter target word that players have to guess. Players have six tries to do so, and attempt to get the word in as few attempts as possible, each time using a valid English word. 

There are 11,881,376 possible 5-letter words if taking every possible sequence of five letters. Even restricting it to words found in English dictionaries and those in common usage today would only drop it down to around 12,000 and 4,000 words respectively\cite{oed}. For a person to randomly guess a target word in six tries would statistically be almost impossible. This, however, is where tile color feedback comes into play. For a given guess word, for each letter Wordle returns a green tile if the corresponding letter is in the target word and in the right location, a yellow tile if the corresponding letter is in the true word but in the wrong location, and a gray tile if neither of these is true. With this information, most players are able to guess the word from thousands of possibilities within six tries. 

The game has captured the attention of millions, with people taking to social media to share their guess results and comment on the difficulty for certain words. One Twitter account that has popped up as a result of this trend is ``@WordleStats'', a bot that tallies all posted Wordle attempts and the distribution of attempts each day. Via this data, we can discover a wealth of information about Wordle player behavior. Particularly, in this paper we develop a model which utilizes both the trends in twitter reporting and the resulting inferred difficulty of target words gleamed from this data to predict future Wordle statistics.  

\section{Data}
\subsection{Data cleaning}

\paragraph{Data errors} We fixed several errors that appear in the provided data by referencing the Twitter posts of the @WordleStats Twitter bot. These are logged below for full transparency.
\begin{itemize}\setlength\itemsep{0.1em}
    \item Day 239: hardmode 3249 $\rightarrow$ 9249
    \item Day 314: tash $\rightarrow$ trash
    \item Day 500: hardmode 3667 $\rightarrow$ 2667
    \item Day 525: clen $\rightarrow$ clean
    \item Day 529: Reported players 2569 $\rightarrow$ 25569
    \item Day 540: naïve $\rightarrow$ naive (ï is not a letter in Wordle)
    \item Day 545: rprobe $\rightarrow$ probe
\end{itemize}

\paragraph{Percentages to counts}
Because the percentages of reports in the different categories are rounded and do not necessarily sum to 100, we divided the percentages in each row by their sum to obtain proportions. As our Bayesian model predicts the \textit{number} of players in each category, we converted these proportions into counts by applying the following method for each row:
\begin{enumerate}\setlength\itemsep{0.1em}
    \item Multiply the proportions by the number of reports on that day to obtain ``counts'' with decimal values
    \item Round the counts down
    \item Add 1 back to the counts, in order from the count which was rounded down most to that which was rounded down least, until the total again matches the number of reports on that day. 
\end{enumerate}
This method gives counts which correspond to the given percentages, are integers, and whose sum is the number of reports on that day.

\subsection{Wordbank}
To model 5-letter words and their properties, we rely on the Stanford GraphBase (SGB) wordbank of 5757 5-letter words created by Donald Knuth \cite{knuth}, which provides a good approximation of the set of words that a player could guess and expect as a target. This word bank is then used in a few different ways. First off, it is used to build the Order Frequency table and the Letter Frequency table. The Letter Frequency table tells us how often each letter appears in 5-letter words, and follows what we would expect. S and E are the most common letters, followed by A and O. The Order Frequency table then shows given that a certain letter is in the word, what proportion of the time is that letter in each position (e.g. given A is in a word, it is the 4th letter $24\%$ of the time).

\begin{figure}[h!]
    \centering
    \includegraphics[width=5cm]{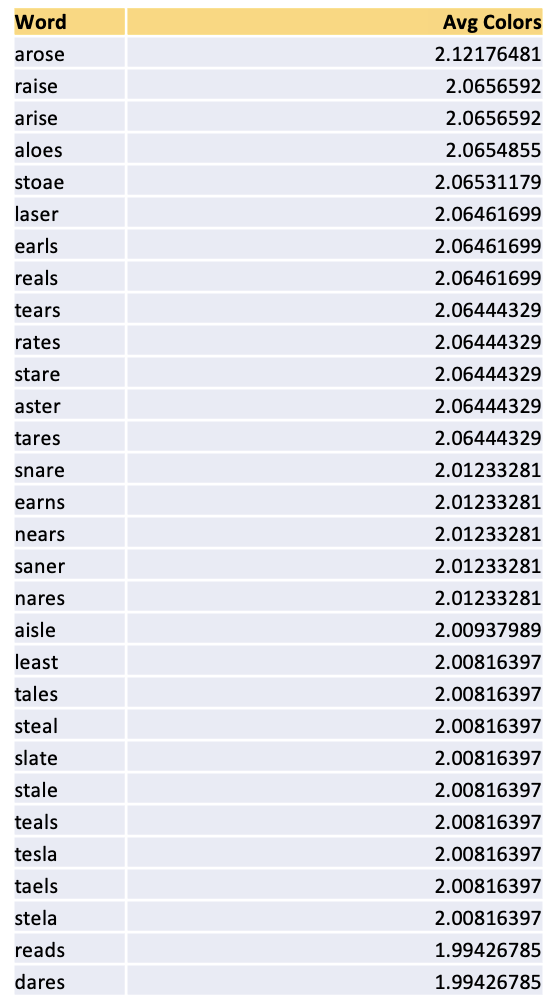}\hspace{5cm}\includegraphics[width=5cm]{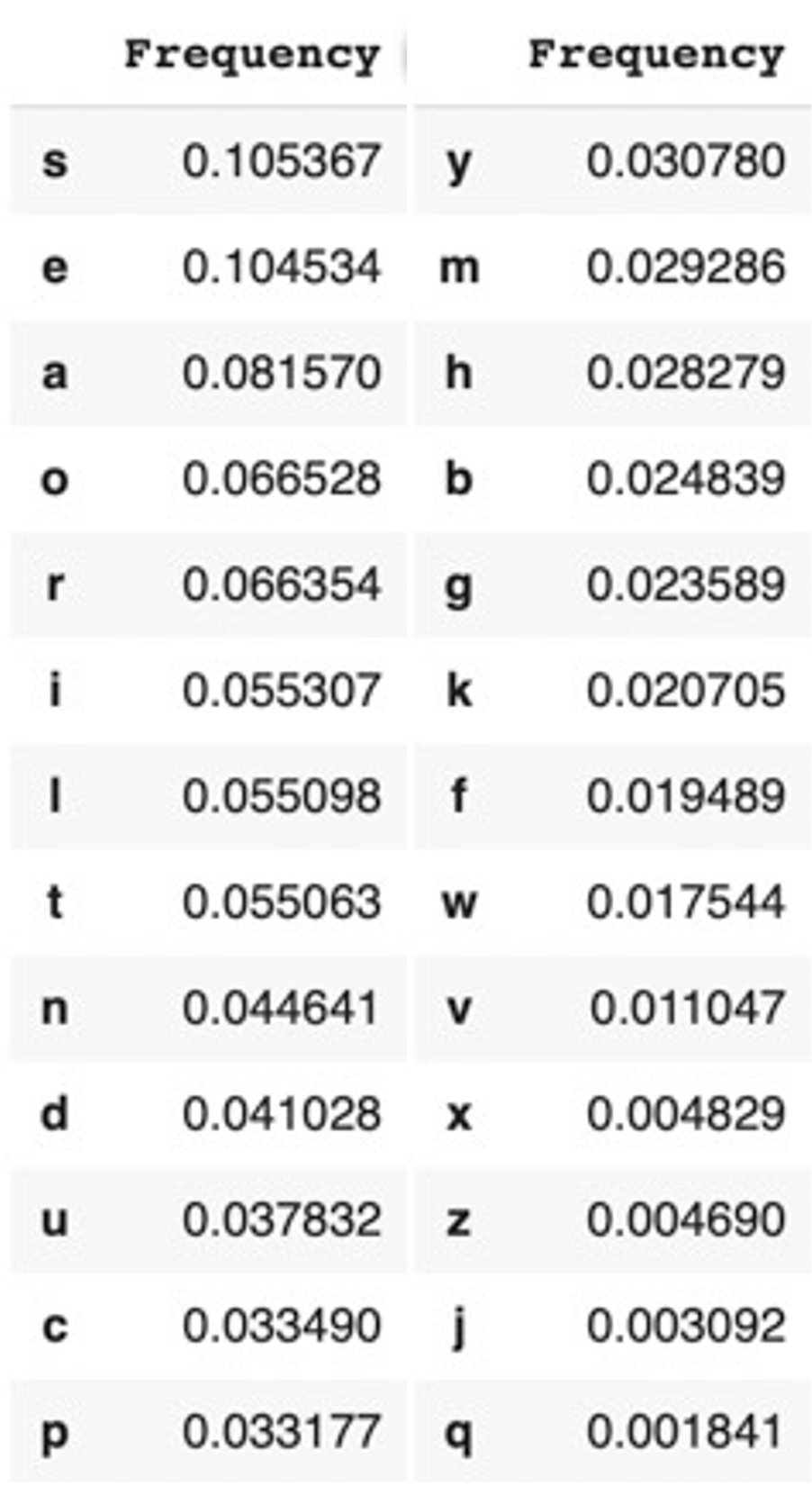}
    \caption{Left: 30 Most common words based on overlap. Right: Frequency of Letters in the SGB wordbank}
    \label{fig:freq}
\end{figure}

As will be detailed later on, we also use this table in numerous word specific calculations, namely computing for a given word $t$ the average number of green, yellow and colored tiles that are returned on any guess chosen uniformly at random from the SGB wordbank when $t$ is the actual target word.  Additionally, for any word $g$ we compute the average number of colors returned when a target word $t$ is chosen uniformly at random from the SGB wordbank and $g$ is guessed, which gives a somewhat naive but reasonable metric to evaluate guess words players would use.  Using this guess word metric, we compile a list of the 30 words with the highest corresponding average, which we take to be a set of common guess words (see Figure \ref{fig:freq}).  This list will be used in the calculation of Subset Entropy later on.

\section{Word \& Difficulty Representation}

Many factors contribute towards the difficulty associated with a given word. For example, \texttt{``zingy''} intuitively seems to be difficult for a variety of reasons - it has uncommon letters (``z'' and ``y''), only one ``canonical'' vowel, and is a generally infrequently used word in English.  A word like \texttt{``onion''} on the other hand would also seem to be difficult, despite all of its letters being fairly common and its usage in every day spoken language being much higher. The reason it is perceived as difficult is due to a repetition of the letter ``o'' and ``n'', which people may be less likely to guess again once they have already discovered one position of.  Thus, given a word, our first task was to list and quantify such characteristics, so that when evaluating word difficulty later on we could instead simply consider the vector of values corresponding to these relevant characteristics. 

\subsection{Vowels}

In all letter-guessing based games (beyond Wordle think hangman), a typical strategy is to exploit the higher frequency of letters which are vowels in words.  In Wordle, the presence of particular vowel should tend to be discovered faster than those of non-vowels, leading to a reasonable assumption that words with more vowels will on average be easier to guess.  Thus, one characteristic computed and considered for each word was the number of vowels it contained (excluding ``y'', as this is a traditionally uncommon letter and hence does not align with the reasoning given above for why we consider vowels in the first place).

\subsection{Usage}

It makes sense that words which are used more commonly will be more familiar to people, and hence easier to guess from given clues.  Using the $\texttt{wordfreq}$ library \cite{wordfreq} in python, this value was easily returned for each word under consideration.

\subsection{Green \& Yellow Tiles}

Another desirable feature of a target word is that there is a high likelihood that after any given guess Wordle will return a large number of green and yellow squares.  Once this occurs, players get very direct hints that they can immediately put into action.  Thus, for any given word, we computed the the average number of green, yellow, and colored (i.e. non-gray) tiles returned considering the given word as the target and assuming guesses were drawn uniformly at random from the SGB wordbank.

\subsection{Unique Letters}
When Wordle returns a color for a letter, it does not tell the player how many times that certain letter appears.  Thus, while it may be easier to initially obtain green tiles for words with repeated letters, players may also be less likely to guess those same letters again in different positions. As a result, it seems justified to consider the number of unique letters in a word as playing a part in the difficulty of guessing it, and hence this value was computed for each considered word. 

\begin{figure}[h!]
    \centering
    \includegraphics[width=5cm]{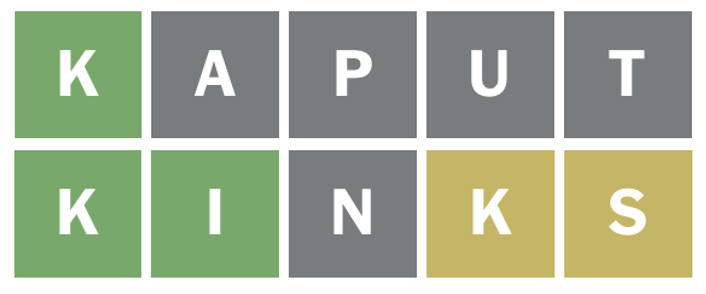}
    \caption{The repeated letter does not show up if the first guess is correct, it gives no indication that there is a second k.}
    \label{fig:repeat}
\end{figure}

\subsection{Entropies}

One of the core ideas in information theory is that of entropy, which is a measure that quantifies the amount of information conveyed by a given event. In the context of a Wordle game, we can consider an event to be a particular choice of a player's move, i.e.~a guess word along with the corresponding positional colors Wordle returns as a result.  Clearly this event gives us some information about the target word, and a reasonable question to ask is how much information?  Once such a value is quantified formally, one can consider the ``best'' guess words to be those which result in a higher amount of information on average (over a uniform distribution of all possible 5-letter target words). Conversely, target words are more difficult to guess which, by some efficient text encoding, contain more information, \textit{or} which result in lower amount of information being obtained on average (over some distribution of possible guessed words). These two ideas idea are formalized via two entropy formulations we give below, the first being a more typical application of the standard Shannon Entropy function, and the second being a more novel notion of entropy we developed specific to Wordle itself.

\subsubsection{Positional Entropy}
The original use case of entropy came from encoding, with more frequent symbols and common arrangements taking less bits to transmit. Our Positional Entropy takes this approach by considering a word to contain more information if its letters and letter arrangements are less common/more unusual. Using the Letter Frequency table and Order frequency table from the SGB word bank, for a given letter $\phi$ we calculate $P_i(\phi)$ as the probability a word has the letter $\phi$ in its $i^\text{th}$ location.  These probabilities can then be plugged into the standard Shannon Entropy function $H$ \cite{shannon_1948}, where $X$ denotes a 5-letter word and $X_i$ corresponds to its $i^\text{th}$ letter, as follows: 
\[H(X) := \sum_{i=1}^{5} -P_i(X_i) \log_2(P_i(X_i))\]
Note that under this metric, words which have more unusual letter placements, and hence can reasonably be assumed to be more difficult, will have a higher value. 

\subsubsection{Subset Entropy}

\begin{figure}
    \centering
    \includegraphics[page=2,width=0.7\textwidth]{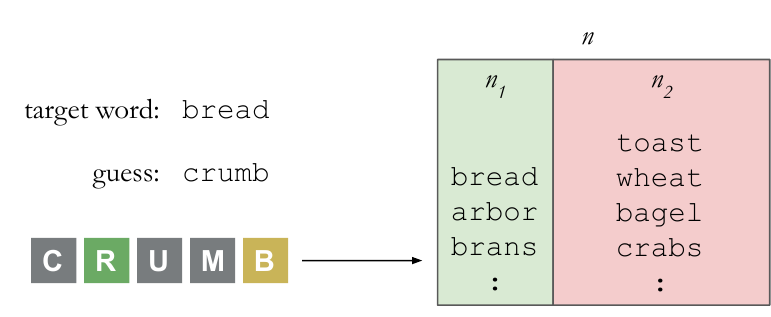}
    \caption{Diagram representing the Subset Entropy method. For a target word \texttt{bread}, a guess \texttt{crumb} is made and Wordle tells the player that \texttt{R} must be in the second position and \texttt{B} is somewhere in the word, but not in the last position. Furthermore, Wordle tells the player that \texttt{C}, \texttt{U} and \texttt{M} are not in the target word. This information defines a subset of $n_1$ words that are compatible with it.}
    \label{fig:steven-entropy}
\end{figure}

Our second entropy formulation is that of Subset Entropy, which is a novel Wordle-inspired metric we developed that, on a given word, quantifies the average amount of information which is revealed about it following one Wordle guess chosen from some distribution.  This metric is motivated by the idea that on a given target word $t$ with a chosen guess word of $g$, information is obtained via the output colors and their corresponding positions which allows one to disqualify other candidate target words $t'$.  For example, if the target word under consideration is \texttt{bread} and the word \texttt{crumb} is guessed, then the corresponding output from Wordle would allow us to eliminate words such as \texttt{toast} and \texttt{crabs} as possible target words, but not allow us to eliminate \texttt{arbor}.  An example illustrating this idea and our notation is given in figure \ref{fig:steven-entropy}.  

Before we detail Subset Entropy, first we define $f_t(g)$ as the factor by which the candidate answer pool shrinks after word $g$ is guessed and $t$ is the target word.  More formally, considering the notation in figure \ref{fig:steven-entropy} we have:

$$f_t(g) = \frac{n}{n_1}$$
where $n$ is the size of the original candidate answer pool and $n_1$ is the size of the resulting candidate answer pool. As an example, a guess $g$ when the given target word is $t$ which results in the possible answer pool shrinking from $4,000$ to $1,000$ words would have $f_t(g) = 4$.  Note that if $g_1$ is a better guess than $g_2$ given the target word is $t$, we have that $f_t(g_1) > f_t(g_2)$, while if a target word $t_1$ is easier to guess than $t_2$ using a guess $g$ then $f_{t_1}(g) > f_{t_2}(g)$.

In typical entropy style, we take the base 2 logarithm of this factor which is computed, and define:

$$I_t(g) = \log_2(f_t(g))$$

We use this value to correspond to our notion of the information conveyed by the event of guessing $g$ on target $t$.  The motivation for choosing this value as opposed to $f_t(g)$ directly is that Subset Entropy will aim to quantify the average information following the first guess when word $t$ is the target, and hence will need to take an average over some quantification of information.  As $f_t$ effectively measures a multiplicative factor of information (rather than additive), taking a geometric rather than arithmetic mean would be more appropriate.  However, as the average of the log value of some terms exactly corresponds to the log of the geometric mean of these terms, we can effectively instead consider the expectation of $I_t(g)$ over some distribution of guesses $\mathcal D$ as a measure of the geometric mean of the factors.  With this justification in hand, we have the following formulation of Subset Entropy:

$$e(t) = \mathbb{E}_{g \sim \mathcal D}[I_t(g)]$$
where $\mathcal D$ is some distribution over the possible guess words. 

In the actual implementation of our model, we considered $\mathcal D$ to be a uniform distribution over the 30 most common guess words list (see figure \ref{fig:freq}).  In particular, this model assumes that players guess one of the 30 best words to guess according to the average color metric.  We decided this was reasonable, as many common words guessed or words similar to them, such as ``slate'', appeared to be on the list.

\subsection{Representation of word difficulty}
\begin{figure}[h!]
    \centering
    \includegraphics[height=0.23\textheight]{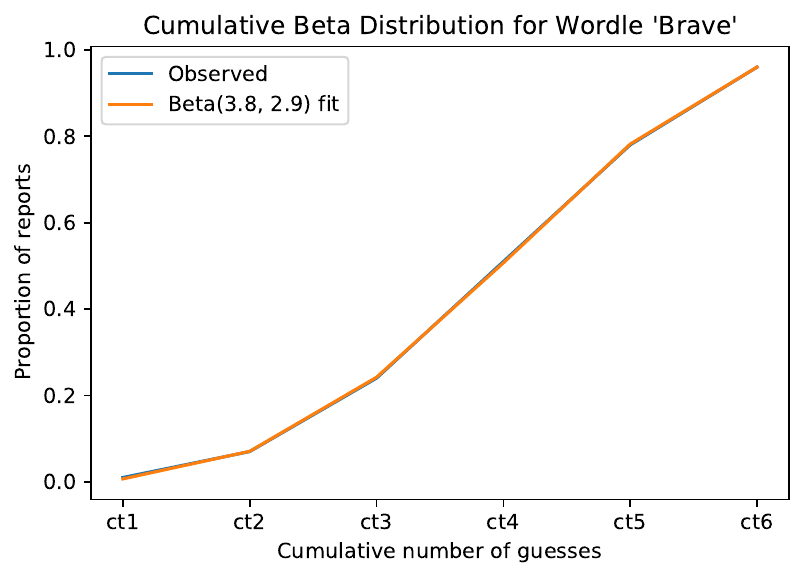}
    \includegraphics[height=0.23\textheight]{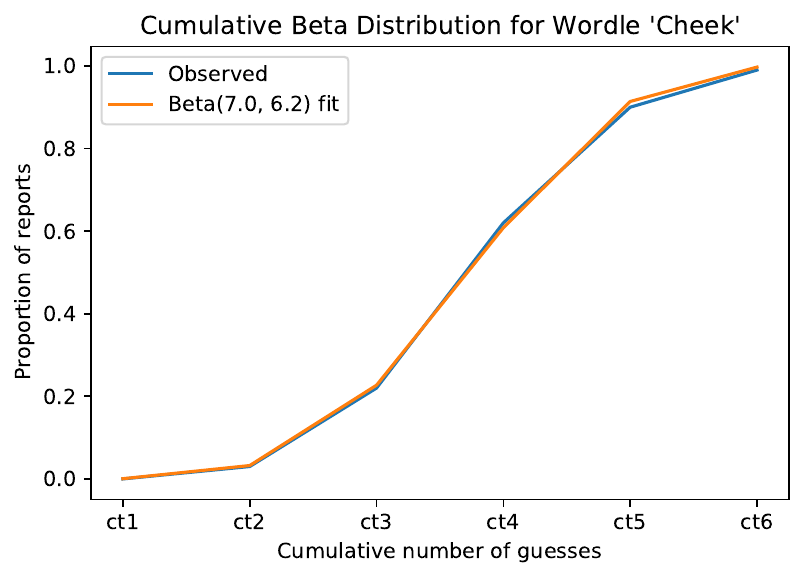}
    \includegraphics[height=0.23\textheight]{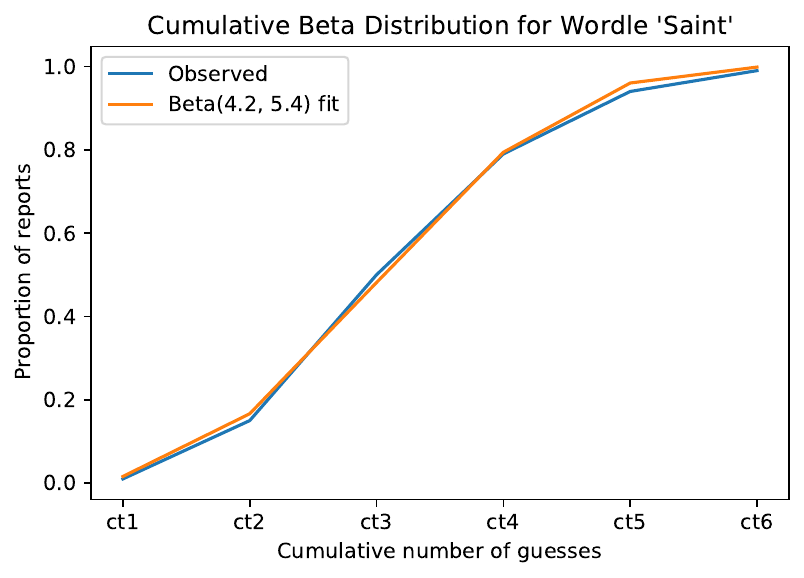}
    \includegraphics[height=0.23\textheight]{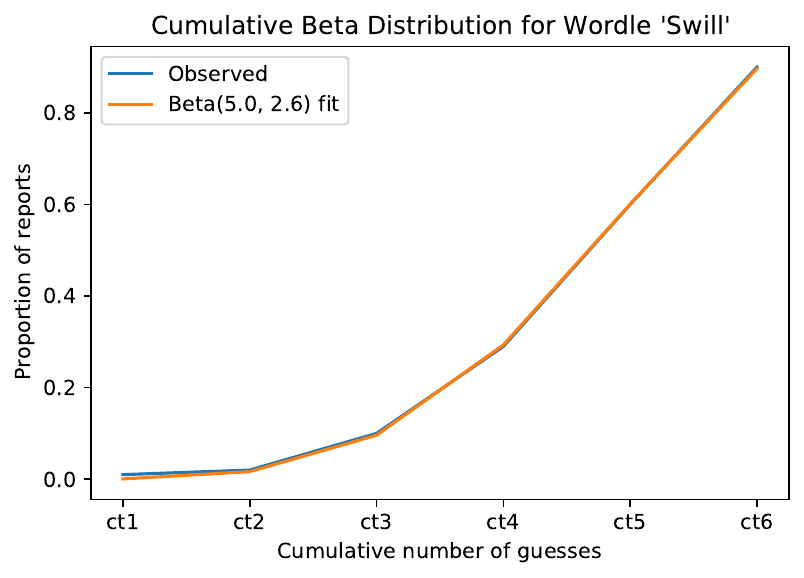}
    \caption{Beta Distribution fit to the cumulative proportion of guesses for Wordle words of various difficulty.}
    \label{fig:beta-dist}
\end{figure}

To aid in modeling and interpretation, ideally the 7-vector data (of players who got the word in 1 try, 2 tries, \ldots, failed) which effectively represents the observed difficulty of a word, should be encapsulated with fewer numbers. We can view these seven categories as a discrete space and the observed proportion as a discrete probability mass function over them. By assuming an ordering of $1<2<\ldots<X$, we then take the cumulative mass function that corresponds to the probability mass function. % we can Another way this data can be interpreted is by looking at 'N-or-less attempts', where it takes the sum of probabilities of everything beforehand. This is then the cumulative distribution function, and we get the equation:
% \[
%    Pr(x \leq N) = \sum_{i=1}^N t_i 
% \]
After embedding the domain into the interval $[0,1]$ by the map ($i$ tries)$\mapsto$ $i/7$ and $X \mapsto 1$, it turns out that the Beta distribution, a two-parameter continuous distribution over $[0,1]$, has a cumulative distribution function that can fit the embedded cumulative mass function closely (see figure \ref{fig:beta-dist}). We then model reader attempts with the cumulative distribution of the Beta distribution, whose probability density function is:
\[
f(x; \alpha,\beta) = \frac{1}{B(\alpha,\beta)}x^{\alpha  -1} (1-x)^{\beta -1},
\]
where $B(\alpha,\beta)$ is the Beta function.
To figure out the exact values of $\alpha, \beta$ that best fit the embedded cumulative mass function, we use non-linear least-squares as implemented in \texttt{scipy}\cite{scipy}.

\begin{figure}
    \centering
    \includegraphics[height=0.3\textheight]{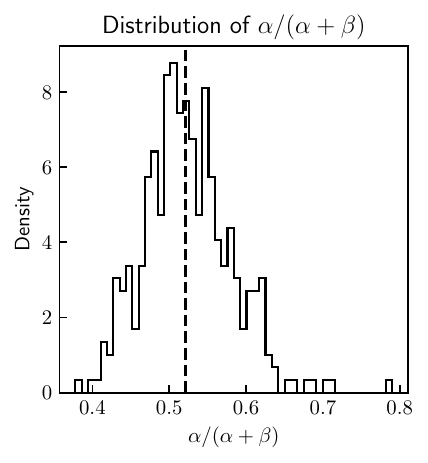}
    \caption{Observed distribution of $\alpha/(\alpha+\beta)$, a measure of word difficulty.}
    \label{fig:alphabeta-dist}
\end{figure}

\noindent By observing the figures above and from the properties of the Beta distribution, we can conclude a few properties of $\alpha$ and $\beta$:
\begin{itemize}
    \item As $\alpha \uparrow$ while all else is held equal, the word is considered more difficult
    \item As $\beta \uparrow$ while all else is held equal, the word is considered easier
    \item $\alpha + \beta$ will give how concentrated the distribution is around the expected value, with higher values being more concentrated.
    \item The ratio $\frac{\alpha}{\alpha + \beta}$ gives the expected value of the Beta distribution, which is a representation of the difficulty of a word, with higher values corresponding to a higher difficulty.
\end{itemize}
Because of these properties, particularly the final one, $\alpha$ and $\beta$ together are an effective and simple representation of the difficulty of a word. Figure \ref{fig:alphabeta-dist} shows $\alpha/(\alpha+\beta)$ computed from the observed reported distributions for all previous words.

% Because of this discovery, rather than predicting the proportion of people for each number of attempts separately, running regression on predicting the $\alpha$ and $\beta$ values will be sufficient. This also ensures that for any prediction, because a beta distribution is used, any parameter that is predicted will have the total proportions sum to 1, and our predictions provides a more intuitive understanding for the difficulty of each word.

\begin{figure}[h!]
    \centering
    \includegraphics[width=15cm]{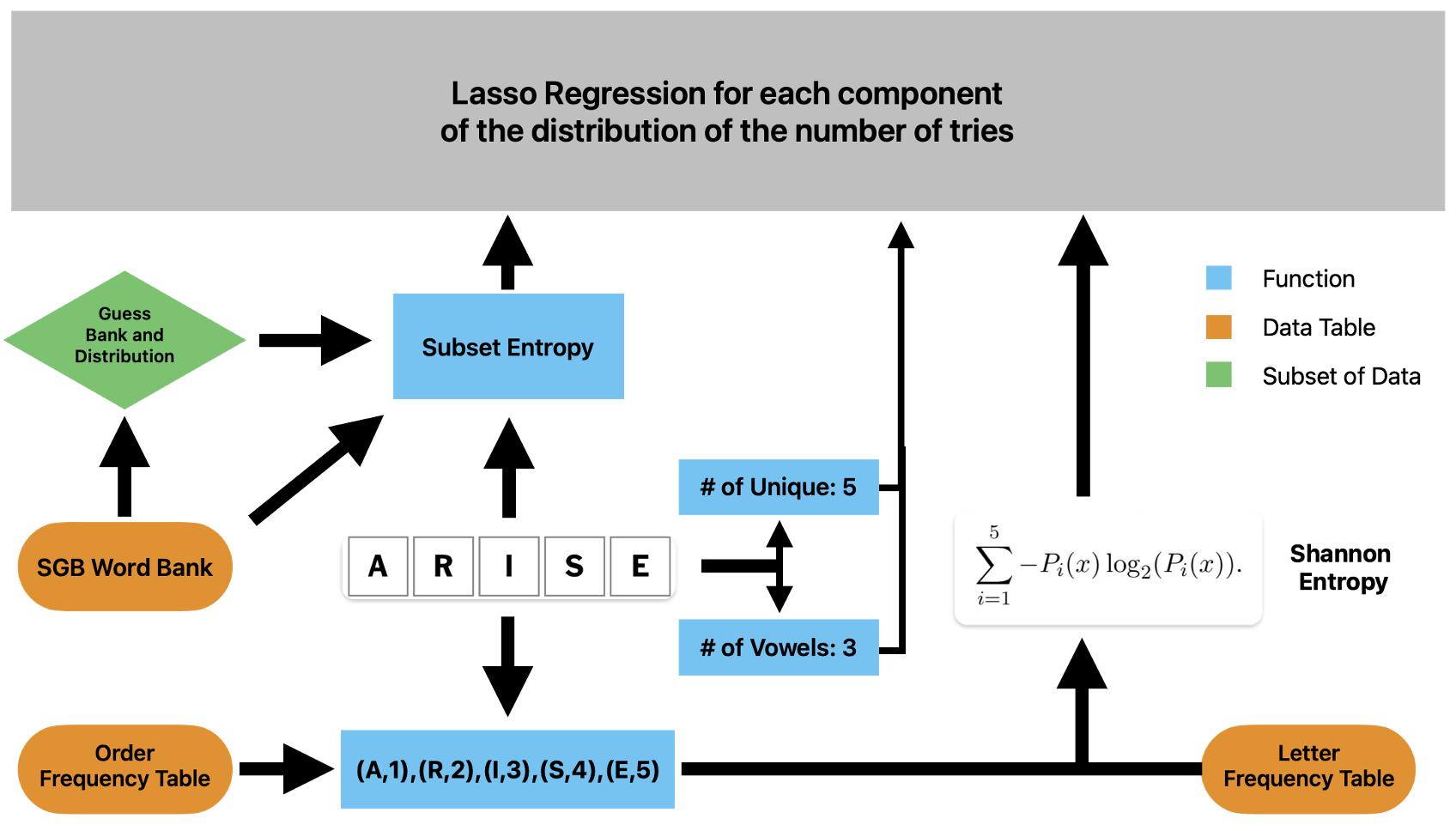}
    \caption{Inputs into the Lasso Regression}
    \label{fig:process-diagram}
\end{figure}

%Each word itself can convey a wealth of information. When Wordle returns a color for a letter, it does not tell the player how many times that certain letter appears, so for words with repeated letters, it may be easier to hit green tiles, but people would also be less likely to assume that the letter appears again. This then means that the number of unique letters should play a part in the difficulty of guessing the word. At the same time, the frequency of letters can also play a part. Certain letters such as 'x' and 'z' are uncommon and hard to guess, but if they return either a yellow or green tile, then suddenly the list of possible words dramatically decreases. 

%These frequencies can then be plugged into the standard Shannon Entropy function: $\sum_{i=1}^{5} -P_i(x) \log_2(P_i(x))$, where $P_i(x)$ is the frequency of letter x in the ith location.

% \begin{figure}
%     \centering
%     \includegraphics[width=0.3\textwidth]{Arise.png}
%     \caption{Frequency of words encoded}
%     \label{fig:entropy}
% \end{figure}

%SGB word database
%https://www-cs-faculty.stanford.edu/~knuth/sgb.html

%Not all yellow and green tiles are created equal

\section{Modeling Methodology}

\subsection{Lasso regression}
Once we have all the variables that can be extracted from the word itself, we must use select a subset of them to include in the Bayesian model, as the time taken by MCMC increases significantly with the number of variables in the model and we must keep the run-time reasonable. To select more important predictors, we use Lasso regression. Lasso regression penalizes large coefficients, which in turn forces less important variables to have coefficients of zero.
Lasso Regression has the following cost function:
\[\min_{\beta} \sum_{i=1}^n \left(y_i - \sum_{j=1}^Px_{ij}\beta_j\right)^2 + \lambda \sum_{j=1}^P |\beta_j|,\]
where $x_{ij}$ is the value of the $j$th predictor for the $i$th data point, and $\beta_j$ is the $j$th coefficient. $\lambda$ is a tuning parameter which after experimentation we set to 0.1 to obtain the desired number of parameters (four). As seen in the cost function, Lasso regression seeks to minimize the error between true and predicted, while also having the sum of all coefficients be small. 

With our parameters, there are seven separate Lasso Regression models being run simultaneously, one on the proportion of people who succeeded in one try, one on those who succeeded in two tries, and so on until the final one for those who failed. Four parameters appeared repeatedly with nonzero coefficients in these regressions: the number of unique letters, the word's usage frequency, the average number of yellow squares revealed, and the Subset Entropy.

\subsection{Bayesian models for prediction}
\begin{figure}
    \centering
    \includegraphics[page=1,width=\textwidth]{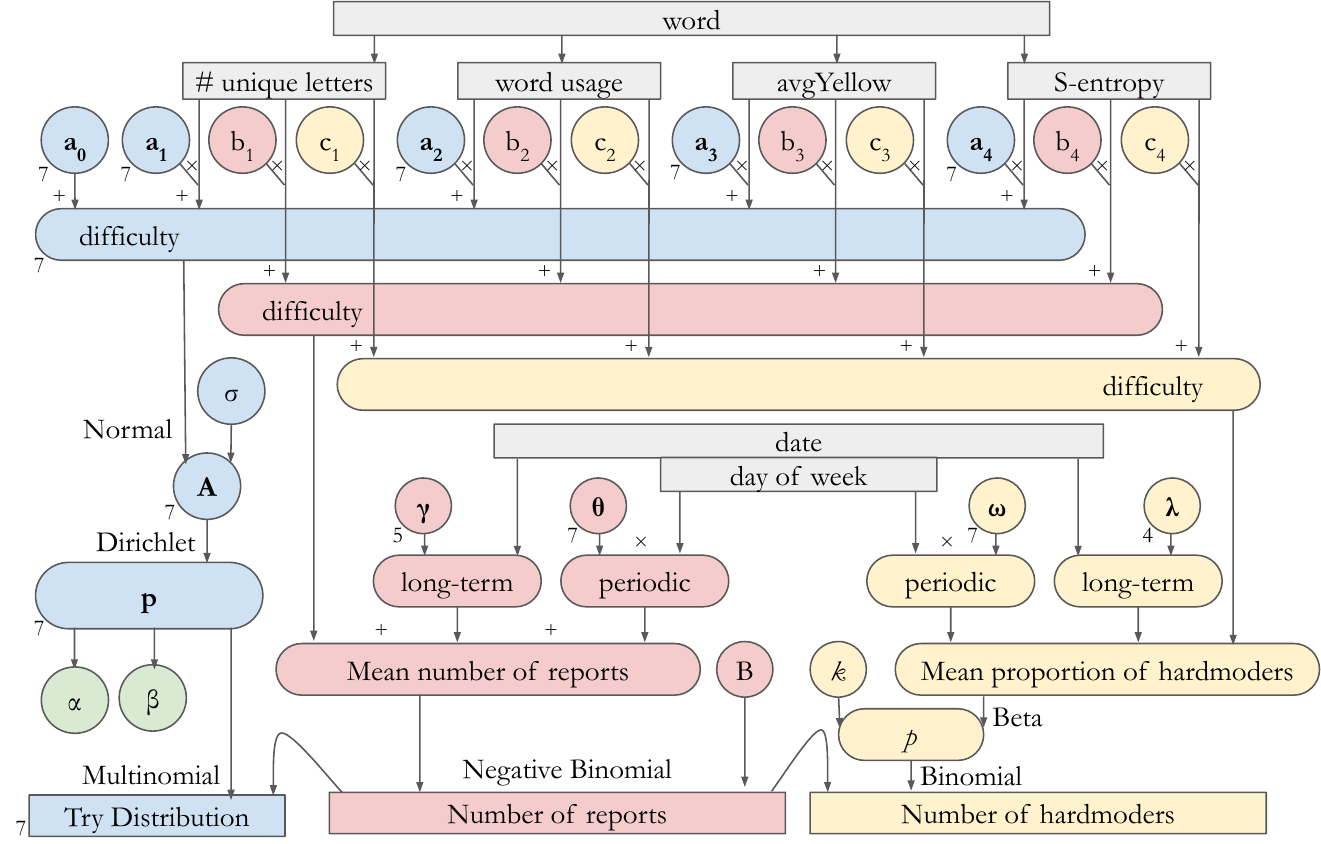}
    \caption{Combined diagram of the three Bayesian models. Shapes with rounded corners are variables, and rectangles are observed data. Numbers indicate the dimension of vectors (all other variables and data are scalars). In yellow is the model for predicting the proportion of hardmoders, in red is the model for predicting the number of reports, and in blue is the model for predicting the distribution of the number of tries. Numbers outside the shapes indicate the dimensions of data and variables when they are not scalars. Conditioned on the observed data, the random variables of all three models are independent, which allows their posteriors to be inferred in three separate runs of MCMC. Note that the number of reports from the Reports model is fed into the Try model and the Hardmoders model. In green are the $\alpha$ and $\beta$ values, which are the parameters of the Beta distribution corresponding to a word's difficulty distribution $\mathbf{p}$.}
    \label{fig:bayes-diagram}
\end{figure}

We use a Bayesian model comprising three conditionally independent submodels to model the distribution of tries (the Try model), the number of reports (the Reports model), and the number of hard-mode players (or ``hardmoders,'' in the Hardmoders model). This model is displayed in Figure \ref{fig:bayes-diagram}.

\subsubsection{Common components}
\paragraph{Word difficulty} Each of the three Bayesian models takes in four predictors that somehow correspond with difficulty. Since the effect of difficulty on the try distribution, the number of reports and the number of hardmoders is not necessarily the same, we do a separate regression within each of these models:
\begin{align}\mathbf d_{\text{try}}&= \mathbf{a_0} + \mathbf{a_1}x_1 + \mathbf{a_2}x_2 + \mathbf{a_3}x_3 + \mathbf{a_4}x_4\label{eq:dtry}\\
d_{\text{reports}} &= b_1x_1 + b_2x_2 + b_3x_3 + b_4x_4\label{eq:dreports}\\
d_{\text{hardmoders}} &= c_1x_1 + c_2x_2 + c_3x_3 + c_4x_4\label{eq:dhardmoders}
\end{align},
where $\mathbf{a}$'s $b$'s, and $c$'s are coefficients to be estimated, and $x_i$'s are predictors.
Note that the Try model has both $\mathbf d_{\text{try}}$ and the $\textbf{a}_i$ coefficients as 7-vectors. Note also that the Try model is the only one with an intercept in this component, as in the others the intercept is captured by other terms. Each of the coefficients has a prior of $\text{Normal}(0,20)$, which is wide enough to be uninformative.

\paragraph{Day-of-week effects}
In both the number of reports and the proportion of hardmoders, there are periodic day-of-week effects. In the Reports and Hardmoders models, these are modeled similarly:
\begin{align}
P_{\text{reports}} &= \sum_{i=1}^7 \theta_i \cdot\mathbf{1}(i\text{th day of week})\label{eq:preports}\\
P_{\text{hardmoders}} &= \sum_{i=1}^7 \omega_i \cdot\mathbf{1}(i\text{th day of week})\label{eq:phardmoders}
\end{align}
The $\theta$ and $\omega$ coefficients are modeled as having $\text{Normal}(0,1)$ prior distributions.

\subsubsection{Try model}
The model for the distribution of tries (the Try model) is fundamentally a Dirichlet-Multinomial model. For a given word on a given day, we model the vector $(t_1, t_2,\ldots,t_7)$, where $t_i$, $i<7$ is the number of reports who got the word on the $i$th try, and $t_7$ is the number of reported failures, as coming from the Multinomial distribution:
\[(t_1,t_2,t_3,t_4,t_5,t_6,t_7)\sim \text{Multinomial}\left(n,7,(p_1,p_2,p_3,p_4,p_5,p_7)\right),\]
where $n$ is the number of reports on that day, and $\mathbf{p} = (p_1,p_2,p_3,p_4,p_5,p_7)$ is a probability 7-vector ($\sum_{i=1}^7 p_i=1$) of the probabilities that a report reports success on the first try, second try, and so on. This vector itself is modeled as coming from the Dirichlet distribution, which is a distribution over discrete probability distributions:
\[(p_1,p_2,p_3,p_4,p_5,p_7)\sim \text{Dirichlet}(7, (A_1,A_2,A_3,A_4,A_5,A_6,A_7)),\]
where $\boldsymbol{A}=(A_1,A_2,A_3,A_4,A_5,A_6,A_7)$ is the vector of concentration parameters for the Dirichlet distribution, $A_i>0$. The logarithm of these concentration parameters is itself drawn from a Normal distribution:
\[\ln(A_i)\sim \text{Normal}(d_{\text{try},i}, \sigma),\]
where $d_{\text{try},i}$ is as defined by Equation \ref{eq:dtry}. $\sigma$ is a variance parameter which has a prior of $\text{Exponential}(1)$.

\subsubsection{Reports model}
The model for the number of reports (the Reports model) is fundamentally an overdispersed Poisson regression. The number of reports on a given day can clearly be modeled as a Poisson distribution with a certain rate. However, since there is additional variation in the observed number of reports beyond just Poisson error that is not accounted for by the predictors, we must use an overdispersed version of the Poisson distribution (i.e. a distribution that behaves like the Poisson, but with additional errors). The Negative Binomial distribution, when parameterized correctly, has this property. %TODO: cite
Hence we model the number of reports for a given word on a given day as:
\[n\sim \text{NegativeBinomial}\left(\exp(L_{\text{reports}}+P_{\text{reports}}+d_{\text{reports}}), B\right).\]
$L_{\text{reports}}$, $P_{\text{reports}}$, and $d_{\text{reports}}$ correspond to the long-term trend in the number of reports, the periodic trend, and the effect of difficulty, respectively. The periodic trend is defined in Equation \ref{eq:preports}. The effect of difficulty is defined in Equation \ref{eq:dreports}. $B$ is a parameter describing the overdispersion, and is modeled as having an $\text{Exponential}(0.01)$ prior distribution so as to allow it to be large.

Because of the correspondence of the long-term trend with the shape of a Gamma distribution, we parameterize $L_{\text{reports}}$ as a function that recalls the probability density function of the Gamma distribution:
\[L_{\text{reports}}=\gamma_1((T-\gamma_5)^{\gamma_{_2}})\exp(-(T-\gamma_5)/\gamma_3)+\gamma_4,\]
where $T$ is a time value for the word. $T$ is scaled such that the first word in the data has $T=0$ and the last word has $T=1$. 
The $\gamma_i$ coefficients have, according to their role in the equation, different prior distributions. $\gamma_1$, $\gamma_2$, and $\gamma_3$ all have an $\text{Exponential}(1)$ prior distribution, $\gamma_4$ has an $\text{Exponential}(0.01)$ prior distribution, and $\gamma_5$ has a $\text{Normal}(0,1)$ prior distribution. These choices of priors are uninformative when considering the scale of $T$.

\subsubsection{Hardmoders model}
The model for the number of people who reported playing in hard-mode (the Hardmoders model) is fundamentally a Beta-Binomial regression. Given the number of reports on a given day, the number of hardmoders is described by a Binomial distribution with a certain probability. However, since there is additional variation in the observed number of hardmoders beyond just the error of the Binomial which is not accounted for by the predictors, the probability itself is modeled as coming from a Beta distribution.

Hence we model the number of hardmoders $n_h$ for a given word on a given day as:
\[n_h\sim \text{Binomial}\left(n,p\right)\]
where $n$ is the number of reports on that day and $p$ is the probability that someone reports playing in hardmode. $p$ itself is modeled as:
\[p\sim \text{Beta}\left(\eta\kappa,(1-\eta)\kappa\right),\]
where:
\[\eta = \text{logit}^{-1}(L_{\text{hardmoders}}+P_{\text{hardmoders}}+d_{\text{hardmoders}})\]
Note that $\eta$ is the mean of this Beta distribution, as $E[p]=\eta \kappa/(\eta \kappa + (1-\eta)\kappa)=\eta$. Correspondingly, $\kappa$ controls the spread of the distribution. 
$L_{\text{hardmoders}}$, $P_{\text{hardmoders}}$, and $d_{\text{hardmoders}}$ correspond to the long-term trend in the number of hardmoders, the periodic trend, and the effect of difficulty, respectively. The periodic trend is defined in Equation \ref{eq:phardmoders}. The effect of difficulty is defined in Equation \ref{eq:dhardmoders}. $\kappa$ effectively controls the overdispersion of the Beta-Binomial model, and is modeled as having an $\text{HalfCauchy}(0.5)$ prior distribution, which is has a long tail and acts as an uninformative prior.

We then parameterize $L_{\text{hardmoders}}$ as a function that can follow the observed shape:
\[L_{\text{hardmoders}}=\lambda_1((T-\lambda_4)^{\lambda_{_2}})+\lambda_3,\]
where $T$ is a time value for the word. $T$ is scaled such that the first word in the data has $T=0$ and the last word has $T=1$. 
The $\lambda_i$ coefficients have, according to their role in the equation, different prior distributions. $\lambda_1$, $\lambda_2$, and $\lambda_3$ all have an $\text{Exponential}(1)$ prior distribution, $\lambda_4$ has a $\text{Normal}(0,1)$ prior distribution. These choices of priors are uninformative when considering the scale of $T$.

\subsubsection{Obtaining the posteriors}
Rather than computing simple point estimates for the parameters, we obtain full posterior distributions given the data. These are obtained by using Markov Chain Monte-Carlo (MCMC) to iteratively sample from the joint posterior distribution of the variables. Because the three submodels are conditionally independent given the data, we run MCMC on the three submodels separately. In particular, we use the No U-Turn Sampler \cite{nuts} as it is implemented in PyMC \cite{pymc}.

\section{Model Results}
Once the posterior distributions of the parameters of three submodels are obtained, we can glean information from the distributions, and both predict and retrodict by feeding words and dates through the model to obtain posterior predictions.
\subsection{Interpretation of parameter posteriors}
The posteriors of the model parameters can tell us something about what attributes of a word affect a word's difficulty, the number of scores reported, or the percentage of hardmoders. The posteriors on the $\mathbf{a}_i$ coefficients obtained in the Try model all indicate an effect on the distribution of results in the seven different categories. A higher number of unique letters, frequency of word usage, average number of yellow squares revealed, or Subset Entropy all cause the word to be easier to guess, which is reflected in positive $\mathbf{a}_i$ coefficients for the number of people guessing the word on the second try, and negative $\mathbf{a}_i$ coefficients for the number of people guessing the word on the sixth try. The word \textit{cause} is used intentionally here. The \textit{correlations} that the coefficients indicate can be interpreted as causal with the simple assumption that Wordle words are chosen completely at random. This is because, if words are chosen randomly, any correlation between a word's properties and the Try distribution must be due to an effect of the word's properties on the Try distribution, and not due to a reverse effect or a confounding variable.

Likewise, the posterior distributions of the $b_i$'s are largely positive. This indicates that, for easier words, more people report their results. The posterior distributions of almost all of the $c_i$ coefficients include zero, suggesting that, beyond the effect of the total number of reports decreasing, there is no additional effect on the number of hardmoders. On the contrary, the posterior for $c_1$ is largely negative, which counteracts the effect of a decreased total number of reports on the number of hardmoder reports. Again, it is possible to read causation from correlation in these results as that is the only possible way to explain the variation under the assumption that a day's Wordle word is chosen randomly. One possible causal interpretation is that the set of players who play on hard mode is more consistent with reporting their results.

Surprisingly, the $\theta_i$ and $\omega_i$ coefficients are centered around zero, indicating no evidence for a periodic effect. This suggests that there are insufficient data to, in combination with the effects of word difficulty and long-term effects, reliably estimate a periodic effect.

\subsection{Retrodiction}
\begin{figure}
    \centering
    Retrodictions for the numbers of reports and hardmoders for past words and dates.
    \includegraphics[height=0.17\textheight]{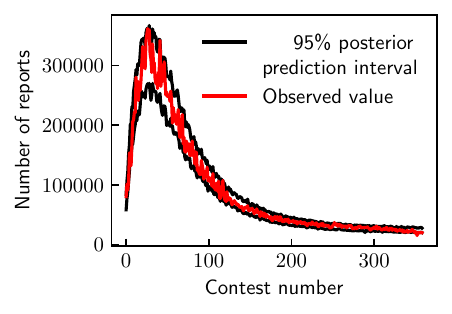}
    \includegraphics[height=0.17\textheight]{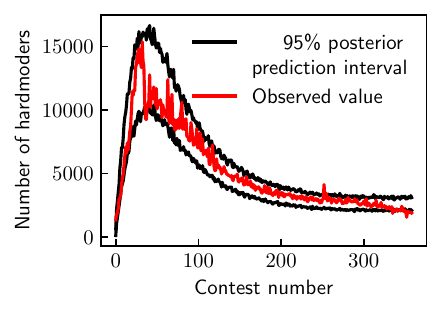}
    \includegraphics[height=0.17\textheight]{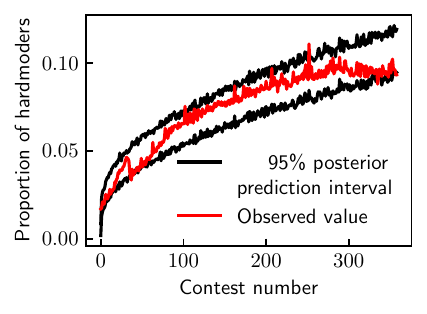}
    \caption{Posterior-predicted distributions for the number of reports, the number of hardmoders, and the proportion of hardmoders over all past words and dates. The models seem to capture the long-term trend and variation in the data well.}
    \label{fig:past}
\end{figure}

\begin{figure}
    \centering
    %Retrodictions for the try distribution of \texttt{fungi}
    \includegraphics[width=\textwidth]{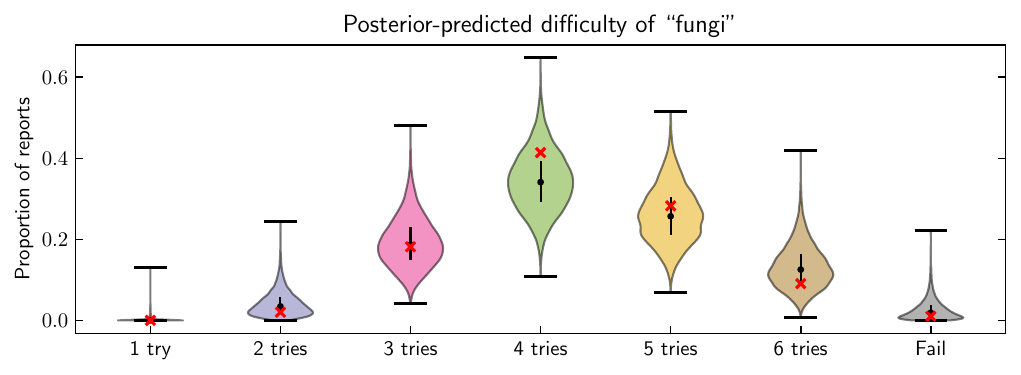}
    \caption{Posterior predictions for the try distribution of \texttt{fungi} as reported by Wordle players. The black points and black lines indicating the median and 50\% interval, and the red crosses indicate the true values.}
    \label{fig:fungi}
\end{figure}

The model can be used to make predictions for past data. Figure \ref{fig:past} shows the posterior predictions for the number of reports, the number of hardmoders, and the proportion of hardmoders for the dates and words given, as well as the observed values. The model seems to capture the variation in the data as well as the long-term trend. Figure \ref{fig:fungi} shows the posterior predictions for the try distribution of \texttt{fungi}, a past Wordle word, and shows good agreement with the truth.
\subsection{Prediction}
\begin{figure}
    \centering
    Posterior distributions for the numbers of reports and hardmoders for \texttt{eerie}.
    \includegraphics[height=0.22\textheight]{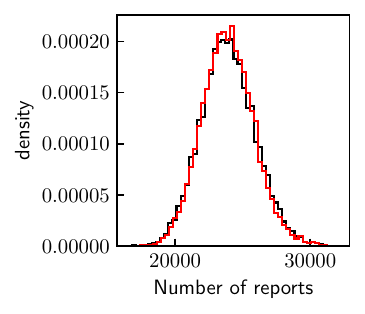}
    \includegraphics[height=0.22\textheight]{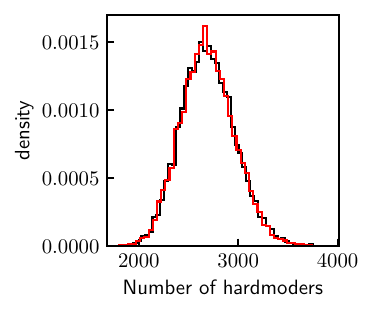}
    \includegraphics[height=0.22\textheight]{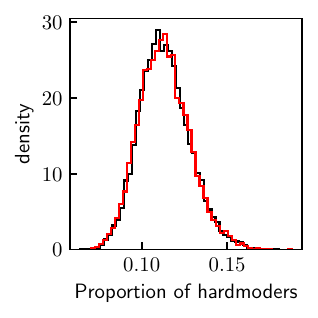}
    \caption{Posterior-predicted distributions for the number of reports, the number of hardmoders, and the proportion of hardmoders for March 1, 2023 (black) and if \texttt{eerie} is the Wordle word on March 1, 2023 (red). For each of the three variables, there is little if any difference between the distributions when averaged over all previous words, and when for \texttt{eerie} specifically.}
    \label{fig:eerie-reports}
\end{figure}

The model can also be used to make predictions for future, unseen data. This can be done either for a future date and word, like \texttt{eerie} on March 1, 2023, or for a date alone (by the posterior predictions for that date averaged over all observed Wordle words). Figure \ref{fig:eerie-difficulty} shows predictions for the difficulty of \texttt{eerie}. Figure \ref{fig:eerie-reports} shows, in general and for \texttt{eerie} in particular, predictions for the number of reported scores on March 1, the number of hardmoders, and the proportion of hardmoders. Table \ref{tab:pred} gives prediction intervals for several different quantities of interest if \texttt{eerie} is the word on March 1.

\subsection{Difficulty representation}
\begin{figure}
    \centering
    \includegraphics[width=\textwidth]{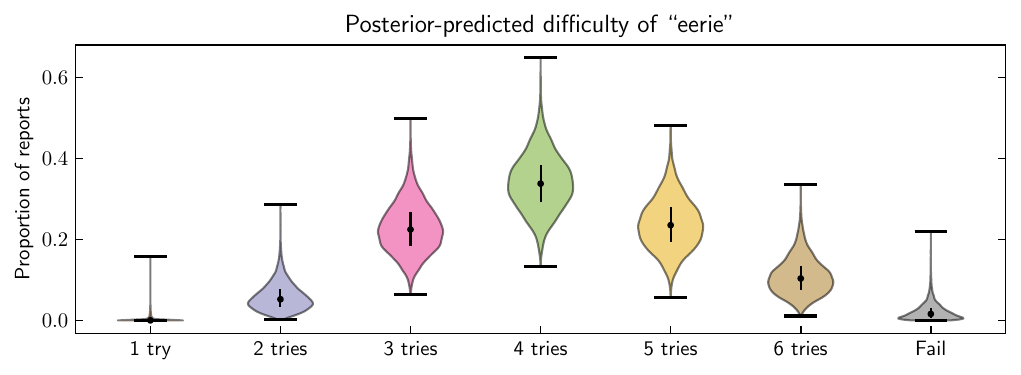}
    \includegraphics[height=0.3\textheight]{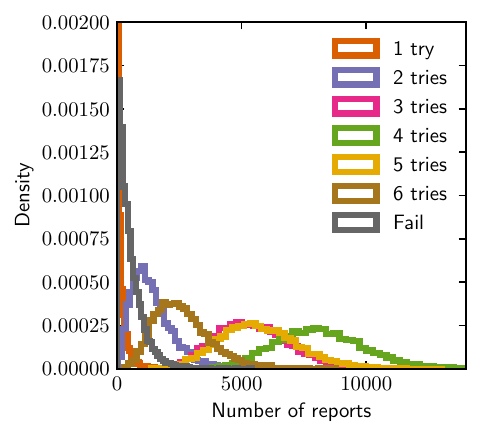}
    \includegraphics[height=0.3\textheight]{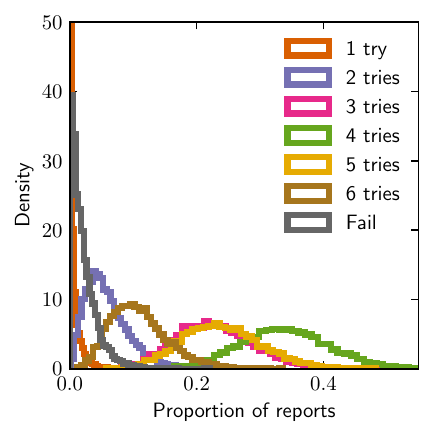}
    \includegraphics[height=0.33\textheight]{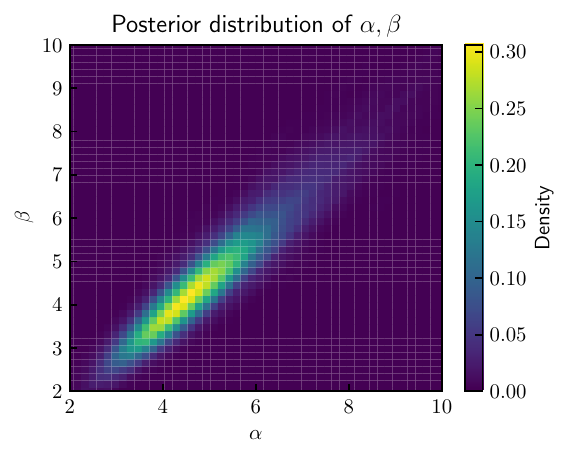}\includegraphics[height=0.33\textheight]{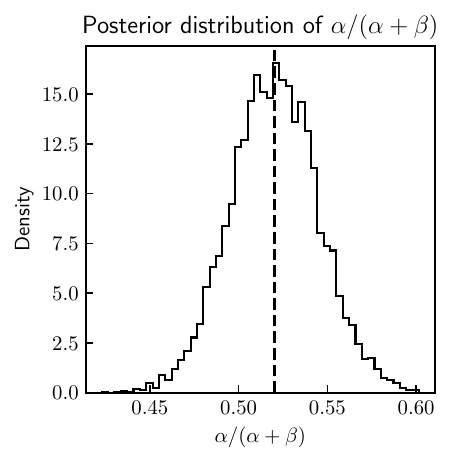}
    \caption{Posterior predictions for several representations of the difficulty of \texttt{eerie} as reported by Wordle players. Top: violin plot of the posteriors on the proportion of reports in each of the seven categories, with the point and black lines indicating the median and 50\% interval. Middle: histograms of the number of proportion of reports in each of the seven categories. Bottom: Posterior distributions of the fitted $\alpha,\beta$ and the expected value of the corresponding Beta distribution, with a dashed line indicating the median. Our model predicts that \texttt{eerie} is a word of middling difficulty, with most players having guessed it on the fourth try or before.}
    \label{fig:eerie-difficulty}
\end{figure}

Given samples from the posterior distribution of a word's $\textbf{p}=(p_1,\ldots,p_7)$ 7-vector representing difficulty, we can use the embedding and optimization method described above to compute posterior samples of the $\alpha$ and $\beta$ values for a word. This is done for $\texttt{eerie}$ in Figure \ref{fig:eerie-difficulty}. Notably, for $\texttt{eerie}$, the posterior samples for $\alpha$ and $\beta$ lie along the $\alpha=\beta$ line, indicating a fairly stable difficulty estimate of $\alpha/(\alpha+\beta)\approx 0.52$ but less certainity about the exact shape of the distribution (i.e. whether most people will get the word in three or four tries, or whether it will be more even between, say, two, three, four, and five tries). This places \texttt{eerie} in roughly the 50th percentile of words when compared to previous Wordle words according to this measure of difficulty.

\begin{table}
    \centering
    \caption{Prediction intervals of several quantities of interest for \texttt{eerie} appearing on March 1, 2023.}
    \begin{tabular}{c|cccc}
        Variable &  95\% & 80\% & 50\% & Median\\\hline
        Number of reports & [20238, 27876] & [21479, 26365] & [22622, 25169] & 23884\\
        Number of hardmoders & [2194, 3239] & [2355, 3048] & [2509, 2870] & 2683\\
        Percentage of hardmoders & [9.97, 12.62] & [10.41,  12.15] & [10.79, 11.72] & 11.25\\
        Percentage in 1 guess & [0, 2.4] & [0, 0.82] & [0, 0.15] & 0\\
        Percentage in 2 guess & [1.09, 14.01] & [2.08, 10.45] & [3.39, 7.70] & 5.23\\
        Percentage in 3 guess & [12.16, 35,85] & [15.34, 31.03] & [18.49, 26.82] & 22.46\\
        Percentage in 4 guess & [21.29, 48.16]& [25.19, 43.2]&  [29.2, 38.51] & 33.79\\
        Percentage in 5 guess & [12.48, 37.09]&[16.16, 32.19]& [19.4, 27.96] & 23.54\\
        Percentage in 6 guess & [ 3.73, 21.33]&  [5.6,  16.99]&  [7.6,  13.53] & 10.37\\
        Percentage failed & [0.06, 7.77]& [0.24, 4.91]& [0.66 3.09] & 1.6\\
    \end{tabular}
    
    \label{tab:pred}
\end{table}

\section{Model Evaluation}
\subsection{Limitations}
\begin{itemize}
    \item The Subset Entropy predictor does not capture how difficult a word is to guess after the first guess. For instance, if a player, after the first guess, has revealed that a word ends in \texttt{atch}, the word could be \texttt{watch}, \texttt{catch}, \texttt{latch}, and so on. Subset Entropy does not reflect this source of difficulty.
    \item Subset Entropy requires us to make assumptions on the distribution of first guesses made by players. 
    \item Compared to traditional statistical methods, Bayesian models sampled using MCMC are extremely slow. Using MCMC to sample from a posterior distribution is not nearly as efficient as using gradient-descent or an analytic solution to find optimal values for the parameters that minimize a cost function
    \item The model does not take into account changes in mean player skill across different days, though time was found to be less relevant than aspects of the word itself in a word's difficulty distribution.
    \item The Reports and Hardmoders models do not take into account autocorrelation in the data and instead assume a stable long-term trend which may underestimate the uncertainty inherent in long-term predictions.
\end{itemize}
\subsection{Strengths}
\begin{itemize}
    \item Using a Bayesian framework allows full distributions to be obtained for any predictions, rather than a simple best-fit and prediction intervals that require approximations and strong distribution assumptions. For instance, other models may not capture the asymmetry in the posterior-predicted distributions for the number of people guessing \texttt{eerie} in 1 try and the number of people failing.
    \item The uncertainty in the Bayesian predictions includes not only that from true noise in the data, but also uncertainty in the estimates of the coefficients and parameters of the model itself.
    \item Our model allows for the parameters of a word to have diverse effects on the Try distribution, the number of reports, and the number of hardmoders, which is appropriate as different aspects of word difficulty may affect these outcomes.
    \item Our model represents the difficulty of a word in just two parameters ($\alpha$ and $\beta$),     which can be predicted with uncertainty. The ``difficulty'' of a Wordle word when played by real people with different backgrounds, strategies, and levels of commitment cannot be exactly measured.
\end{itemize}

\section{Conclusion}
We model how difficult words are to guess in Wordle using several approaches. Rather than having a single, prescriptive metric, we come up with several which correspond to a word's difficulty in different ways. We then select a subset of these predictors using a Lasso regression.

By using the Beta distribution as a representation of distributions of numbers of guesses, we are able to condense the information about the difficulty of a word into just two numbers, and from there into a single number which allows us to directly compare words against each other.

Using an innovative Bayesian model, we are able to predict how many guesses future players will report making on future words, how many players will report their results on Twitter, and how many players will report playing in hard-mode. The innovative aspect of the model is that it is made up of three submodels which, when conditioned upon the data, are independent. Hence the overall model can be quite large while still being efficient enough to perform Markov Chain Monte-Carlo on (as MCMC can be run separately on each submodel) and obtain samples from the posterior distribution. Given the Bayesian nature of the model, it is also able to provide uncertainties on any predictions and retrodictions made in the form of probability distributions instead of simple prediction intervals. We provide several predictions supposing that \texttt{eerie} is the Wordle word on March 1, 2023 and provide several retrodictions to confirm that our model aligns with past data.

%\vspace{5em}
\newpage
\printbibliography

@misc{oed, title={Home : Oxford English Dictionary}, url={https://www.oed.com/browsedictionary}, journal={www.oed.com}, author={Oxford University Press} }

@article{shannon_1948, title={A Mathematical Theory of Communication}, volume={27}, DOI={https://doi.org/10.1002/j.1538-7305.1948.tb00917.x}, number={4}, journal={Bell System Technical Journal}, author={Shannon, Claude}, year={1948}, month={Oct}, pages={623–656} }

@misc{knuth, title={Knuth: The Stanford GraphBase}, url={https://www-cs-faculty.stanford.edu/~knuth/sgb.html}, journal={www-cs-faculty.stanford.edu}, author={Stanford University} }

@software{wordfreq,
  author       = {Robyn Speer},
  title        = {rspeer/wordfreq: v3.0},
  month        = sep,
  year         = 2022,
  publisher    = {Zenodo},
  version      = {v3.0.2},
  doi          = {10.5281/zenodo.7199437},
  url          = {https://doi.org/10.5281/zenodo.7199437}
}

@ARTICLE{scipy,
  author  = {Virtanen, Pauli and Gommers, Ralf and Oliphant, Travis E. and
            Haberland, Matt and Reddy, Tyler and Cournapeau, David and
            Burovski, Evgeni and Peterson, Pearu and Weckesser, Warren and
            Bright, Jonathan and {van der Walt}, St{\'e}fan J. and
            Brett, Matthew and Wilson, Joshua and Millman, K. Jarrod and
            Mayorov, Nikolay and Nelson, Andrew R. J. and Jones, Eric and
            Kern, Robert and Larson, Eric and Carey, C J and
            Polat, {\.I}lhan and Feng, Yu and Moore, Eric W. and
            {VanderPlas}, Jake and Laxalde, Denis and Perktold, Josef and
            Cimrman, Robert and Henriksen, Ian and Quintero, E. A. and
            Harris, Charles R. and Archibald, Anne M. and
            Ribeiro, Ant{\^o}nio H. and Pedregosa, Fabian and
            {van Mulbregt}, Paul and {SciPy 1.0 Contributors}},
  title   = {{{SciPy} 1.0: Fundamental Algorithms for Scientific
            Computing in Python}},
  journal = {Nature Methods},
  year    = {2020},
  volume  = {17},
  pages   = {261--272},
  adsurl  = {https://rdcu.be/b08Wh},
  doi     = {10.1038/s41592-019-0686-2},
}

@article{pymc,
  doi = {10.7717/peerj-cs.55},
  url = {https://doi.org/10.7717/peerj-cs.55},
  year  = {2016},
  month = {4},
  publisher = {{PeerJ}},
  volume = {2},
  pages = {e55},
  author = {John Salvatier and Thomas V. Wiecki and Christopher Fonnesbeck},
  title = {Probabilistic programming in Python using {PyMC}3},
  journal = {{PeerJ} Computer Science}
}

@article{nuts,
  title={The No-U-Turn sampler: adaptively setting path lengths in Hamiltonian Monte Carlo.},
  author={Hoffman, Matthew D and Gelman, Andrew and others},
  journal={J. Mach. Learn. Res.},
  volume={15},
  number={1},
  pages={1593--1623},
  year={2014}
}

\end{document}